# Design of a programmable spiral phase mirror

Fabio Antonio Bovino,[1] Matteo Braccini,[2] and Mario Bertolotti[2]

[1]*Quantum Optics Lab, SELEX Sistemi Integrati, Via G.Puccini 2, 16154, Genova, ITALY*
[2]*Dipartimento di Scienze di Base ed Applicate per l'Ingegneria, Sapienza Università di Roma, Via Scarpa 16, 00161, Roma, ITALY*

We designed a metallic programmable mirror for optical wavelengths made up of a large number of discrete steps which can be moved independently by actuators. The device is able to easily reshape an incident beam, imprinting on it an optical vortex with tunable topological charge and orbital angular momentum (OAM). Furthermore it is possible to generate almost arbitrary field profiles which can be used for spatial encoding.

## 1. Introduction

Since many years singular optics has become an attractive branch of study because of the large number of fields in which its most studied phenomena, optical vortices (OVs), find applications. As a matter of fact the so called "twisted light" of vortex beams was thoroughly studied[1-3] and OVs were employed as optical tweezers for particles manipulation [4], solitons[5], vortex algebra and quantum information[6]. For these purposes the behavior of such beams was also studied in nonlinear phenomena of second harmonic generation [7], difference frequency generation [8] and parametric down-conversion.
As the name singular optics suggests, optical vortices are phase singularities, i.e. are points of the space where the phase of the electromagnetic field is indeterminate and, because of the continuous nature of the field itself, the amplitude vanishes; in particular, beams carrying OVs show helical phase fronts whirling around the vortex axis, which shows itself as a dark core in the beam's profile. That explains why, for defining them, the crystallographic term, screw phase dislocations was used [2]. The main feature of OV beams is their topological charge Q, defined as the integer number of twists of the phase around the singularity's axis. The latter can be evaluated by means of a circulation integral which encloses all the vortices of the beam:

$$Q = \frac{1}{2\pi} \oint d\phi \qquad (1)$$

As pointed out before the phase fronts of a vortex beam are helicoids; this means that the corresponding electric field has a purely azimuthal phase dependence which varies as exp[iQθ], being θ the azimuthal cylindrical coordinate. Examples of vortex beams are the well-known Laguerre-Gaussian and Bessel-Gaussian beams [9].

The helical phase dependence of these kind of beams implies the presence of orbital angular momentum (OAM) which manifests itself through a Poynting vector rotating around the singularity's axis [10]. Indeed, since the angular momentum density is defined as **rxP**, being **r** the transverse position coordinate and **P** the linear momentum density, simple algebraic passages show that OAM is tightly bound to the tangential component of the Poynting vector, thus to the azimuthal dependence of the electric field [11].

In the past years several techniques were proposed to generate beams with OAM; among those should be mentioned the cylindrical lenses based mode converter [12] which converts a Hermite – Gaussian mode into a Laguerre – Gaussian one, the well-known computer generated "fork hologram" (CGH) which impresses the OV through a fork defect in its fringe pattern, and the spiral phase plate (SPP), a transparent device which has a thickness varying linearly with the azimuth angle θ [13]. The latter forces the incident beam to experience an azimuthally dependent phase change, thus creating the helical wavefront. The main flaw of these devices is the lack of tunability in terms of topological charge, operating wavelength and OAM. Adjustable spiral phase plates made of plexiglass were proposed [14] but they have a limit bound to the breaking point of the twisted plate. Better tunability can be reached using spatial light modulators (SLM), however, since the most common are liquid crystals based, the incident beam should not carry excessive power.

An alternative way of generating OVs can pass through the helical shaping of the optical path of a reflected beam. While SPPs work on transmitted beams, a spiral phase mirror can be designed to impose the linear azimuthal phase dependence on the reflected beam. Adaptive helical mirrors were designed and produced [15-16] through a cut in a flat mirror; the latter is then spirally deformed by a tubular actuator. However because of the continuous nature of these devices, these mirrors can only be tuned up to a topological charge Q = 4. Furthermore they limit themselves to generate a spiral ramp and cannot shape arbitrarily the beam spatial distribution and orbital angular momentum. Segmented mirrors made of 37 exagonal cells have also been proposed [17], allowing a good control of the topological charge even though, since is required a purely azimuthal phase profile, the chosen geometry may not be optimal.

For this sake, in what follows we propose a helical mirror composed by discrete steps which can be moved independently by actuators; it is possible to configure the steps as helical ramps to generate optical vortices or as azimuthally dependent profiles which can shape the beam's intensity distribution for spatial encoding.

In the following sections we discuss the design of the helical mirror and the possible experimental setup for using it, then we show numerical simulations of the main configurations of the mirror' steps and the corresponding beam's profiles; finally we draw our conclusions and point out the possible advances.

## 2. Design of the helical mirror

Our discrete deformable device is composed by a number N of azimuthally distributed segments of metallic mirror which can be independently displaced from their original position by actuators as it is shown in Figure 1.

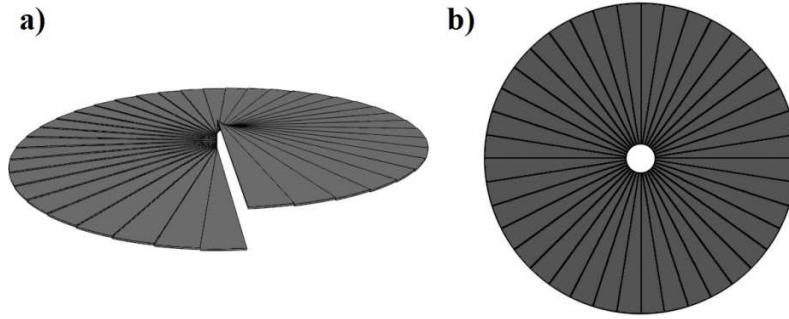

Figure 1. Schematic representation of the azimuthally segmented helical mirror composed by N = 40 steps. a) Perspective view: the steps are displaced in helical configuration; b) top view.

The behavior of the deformable mirror can be treated considering a simple reflection model. In fact when an incident beam is reflected by the mirror it undergoes only to a phase shift due to the total deformation of the devices, while its amplitude remains unaffected. Since the deformation of the mirror surface can only be in the z direction, i.e. the propagation axis, and varies for each step, thus with the azimutal coordinate, a displacement function $\Delta z(\theta)$ can be introduced to take into account the total phase change experienced by the reflected field. Therefore if a beam with amplitude E(r, θ, z) impinges on the mirror with normal incidence, the field resulting from the reflection by the mirror is

$$E_R(r,\theta,z) = E(r,\theta,z)e^{ik\Delta z(\theta)} \qquad (2)$$

being *k* the beam's propagation constant. Looking at Eq. 2 it is evident that it is possible to configure the mirror in order to define an arbitrary height deformation varying along the azimuth direction.

If the composing steps are arranged in a staircase fashion, as happens for the spiral phase plates, the height of the mirror grows linearly with the angle **θ**, thus creating a helical path. This configuration can be used for the generation of optical vortices. As a matter of fact we pointed out before that an OV beam has a phase distribution which goes as exp(*i*Q **θ**); this means that, in order to obtain a vortex of topological charge Q, the phase change along **θ** should be Q times 2π. A simple formula for the mirror profile Δz which links the height of each step to the operating wavelength λ and the topological charge can be found by comparing the phase term of Eq. 2 with the expression of the helical wave fronts

$$\Delta z(\theta) = \lambda Q \frac{\theta}{2\pi} \qquad (3)$$

which is obviously a helical ramp. Since the steps are distributed along θ this direction can be discretized and, by associating each step with an increasing integer index dependent on the θ coordinate, Eq. (3) can be easily written in terms of the step index *j* which range from 1 to N.
However such a profile may result difficult to obtain when considering higher topological charges, thus limiting the operability of the device, as happened for the adaptive helical mirror of ref [15]. A simpler way to generate large topological charges, provided that the mirror is composed by a suitable number of steps N [18], is to configure the device as a saw tooth profile with a number of periods equal to the desired topological charge. In this way each ramp imprints an azimuthally dependent phase shift of 2π reducing drastically the maximum displacement of the steps, therefore overcoming the problems associated with common adjustable helical devices. Furthermore the smaller displacement distances may even reduce the response time of the mirror. Figure 2 shows how the steps can be configured in this situation reducing the problem to the simple case of unitary topological charge.

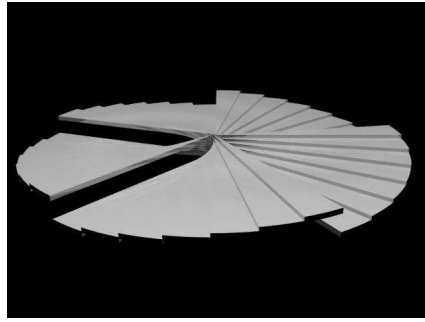

Figure 2. Helical mirror configured in the saw tooth profile with Q = 5; the steps are arranged in five linear ramp along the azimuth coordinate, introducing an equal number of 2π shift.

The helical mirror proposed by Ghai suffered of problems regarding the inversion of the sign of topological charge because of hysteresis issues of the PZT actuator. In a discrete scheme it is sufficient to invert the direction of the ramps to change the sign of the topological charge.
Of course the independency of the steps allows to program the mirror with an arbitrary azimuthal profile in order to generate different space profiles. For example, by grouping steps together, it is possible to reduce the discretization of the helical ramp, going below the minimum step number suitable to represent a fixable topological charge, thus tuning the OAM. The height profile Δz(θ) can assume also square wave, triangle wave or trapezoidal wave profiles: these are all ways to tune both the angular momentum and the intensity distribution which can be used for spatial encoding.

## 3. Numerical simulations

In order to estimate the features of the designed mirror we let a Gaussian beam hit the mirror, that is it acquires the phase dependence expressed by Eq. (2), and let it propagate in free space. We then consider the intensity distribution in the far field region and evaluate the field's orbital angular momentum.
Supposing to hit the mirror in the waist of the incident Gaussian beam, the diffracted beam on the back focal plane of a lens of focal length $f$ will be proportional to the Fourier transform of the incident field times the phase function exp[iΔz(θ)]:

$$E_R(x,y,z) \propto \iint e^{-\frac{x^2+y^2}{w^2}} e^{ik\Delta z(\xi,\eta)} e^{-i\frac{k}{f}(\xi x+\eta y)} d\xi d\eta \qquad (4)$$

where ξ and η are the transverse Cartesian coordinates in the plane of the mirror. It should be noted that we rewrote the θ dependence of Δz in terms of the coordinates (ξ,η) so that numerical evaluation of Eq. (4) results simpler.

On the Fourier plane we therefore calculate the OAM per photon in ℏ units of the diffracted field as the ratio of the total angular momentum and the field's energy[ ]:

$$\ell = \frac{\int \mathbf{r} \times \mathbf{P} d\mathbf{r}}{\int |E_R(\mathbf{r})|^2 d\mathbf{r}} \qquad (5)$$

being **P** the linear momentum density vector associated with the field's distribution.

For this purpose we considered a mirror composed by N = 360 steps so that a high discretization degree is achieved, allowing to test both large topological charges and to accurately reproduce more or less complex height profiles.

Several mirror's configurations were tested by holding the topological charge and varying the number of steps and by modulating the height profile Δz. In the following we report both the configuration of the step and the field's intensity distribution in the far field for every considered case at an operative wavelength of 830 nm.

*Helical profile*

A topological charge Q = 1 has been chosen and the N steps were grouped together with the same displacement in order to assemble larger sector arranged in a well-known helical distribution. Figure 3 shows the phase profile imposed by the mirror on the incident field together with the diffracted field's intensity for different numbers of sectors N' composing the helical staircase. It can be seen that a suitable $LG_{10}$ like distribution appears for values of N' > 16. In this situation the angular momentum grows approaching unity.

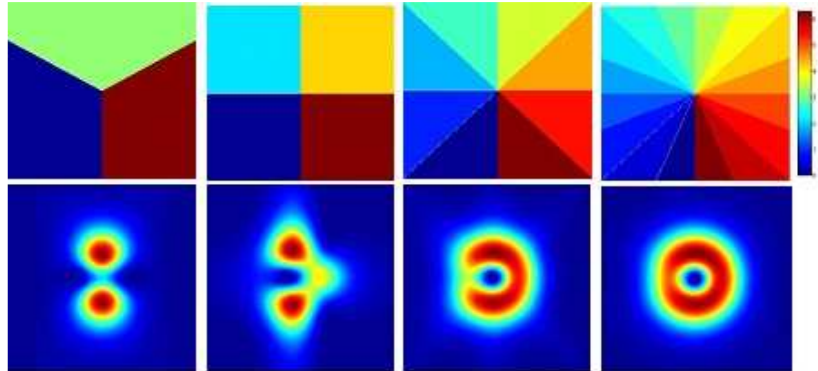

Figure 3: Phase distributions imposed by the mirror (upper row) and corresponding far field intensities (bottom row) for growing number of steps: a) N'= 3; b) N'=4; c) N'=8; d) N'= 16.

*Saw tooth profile*

As we pointed out before, helical wave fronts are better obtained by assembling Q linear azimuthal 2π phase ramps instead of a single staircase with a displacement of 2πQ. In this situation the maximum height difference should be fixed at Δz = λ for each ramp; therefore the unitary ramp so created must be replicated Q times in the whole azimuth distribution. This corresponds, along the coordinate θ, to a saw tooth profile where the number of periods corresponds to the total topological charge.

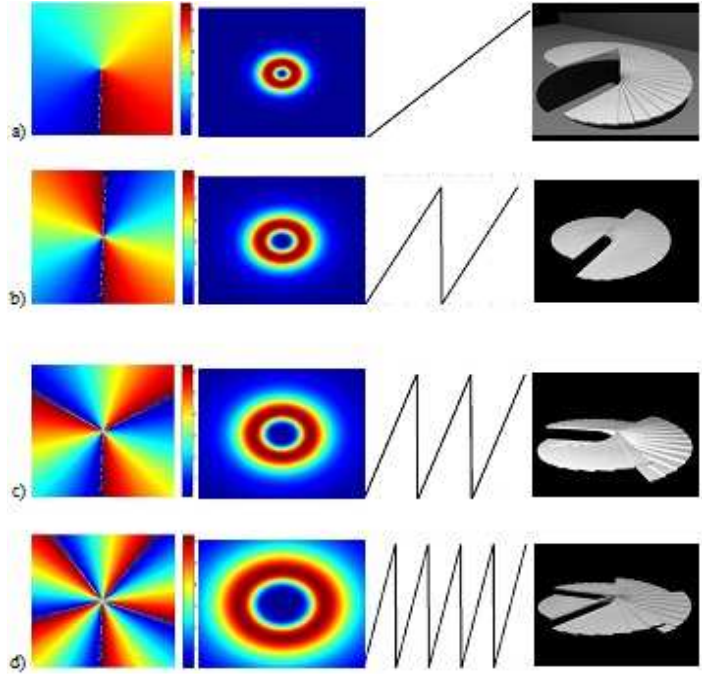

Figure 4: Helical mirror configured in a saw tooth profile; phase profile are reported (first column) together with the corresponding diffracted intensities (second column), the displacement Δz as a function of the azimuth coordinate (third column) and the distribution of the mirror's steps. a)Q = 1, OAM =1; b)Q = 2, OAM =2; c) Q = 3, OAM =3; d) Q= 5, OAM = 5.

In figure 4 the mirror displacement profiles and the corresponding generated intensities and phases are reported for topological charge values up to 5. We evaluated the OAM by using Eq. (5) and found that $\ell = Q$. Obviously the condition on the step number mentioned in the simple helical case remains valid but should be considered for each single ramp.

*Triangular wave profile*
In this case the steps are distributed in a triangular wave, i.e. the displacement grows linearly until it reaches a phase shift of $2\pi$ and then decreases to zero with a negative slope *m*. In our simulations, we fixed the positive ramp' slope and chose the maximum displacement in order to obtain a $2\pi$ phase shift, therefore we changed the steepness of the negative ramp from 1, i.e. a symmetric wave, to infinite, which corresponds to a saw tooth profile. In other terms, we considered the previous case of a saw tooth profile with a fixed topological charge Q and reduced the negative slope until it reaches a symmetric distribution along the azimuth coordinate. This means that while the slope becomes less steep, the number of periods decreases, as can be seen from figure 5.
As we move toward the symmetric case, the phase jump which separates the ramps is gradually removed. Therefore the positive phase accumulated in a fixed azimuthal arc is compensated by the descending ramp. The exact compensation happens when the ramps have the same slope, leading to a zero topological charge, thus a zero orbital angular momentum. In all the intermediate cases the resulting OAM is proportional to the net difference over the circumference of the positive and negative phase contributions.

Figure 5 reports the phase profiles imprinted by the mirror and the corresponding intensity distributions of the reflected field for increasing slopes (modulus) of the descending ramps. It can be observed that the steeper is the slope, the higher is the OAM until it reaches its maximum value which equals the number of periods in the saw tooth case.

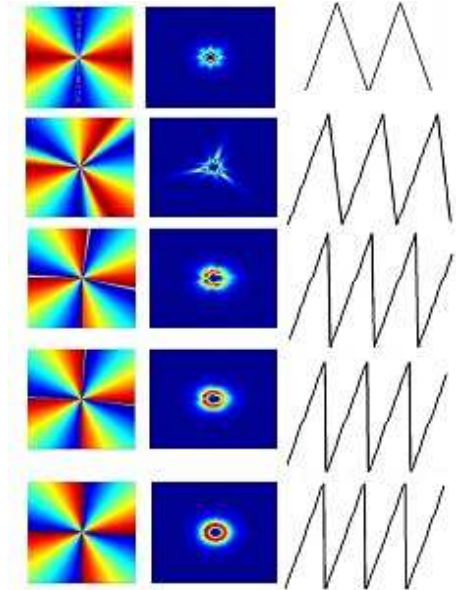

Figure 5. Triangular wave profile with different slopes; the maximum topological charge, i.e. corresponding to saw tooth case is Q = 4. From left to right: imprinted phase profile, far field intensity of the reflected beam, steps' displacement. a) $m = 1$, OAM = 0; b) $m = 3$, OAM = 0.5; c) $m = 20$, OAM = 2.23; d) $m = 40$, OAM = 3.52; e) $m = 100$, OAM = 4.

Looking at figure 5 it can be noticed that these configurations can be thus used in order to perform a tuning of the orbital angular momentum simply controlled by the slopes of the azimuthal profile. Therefore, keeping the maximum topological charge fixed, it is possible to tune the OAM by changing the slope of the ramps.

Another possibility is to keep fixed the number of ramps, i.e. the number of periods, and varying the slope in order to perform an alternative tuning of the OAM. In figure 6 are plotted the phase distributions together with the corresponding intensity far field patterns for a triangular wave with four periods. How was pointed out in figure 5 the increasing slope leads to an increasing OAM which tends to assume the integer value of the topological charge, i.e. the number of periods, in the limit case of infinite slope.

An analogous procedure can be followed with fixed OAM in order to obtain different controllable spatial field's distribution. As a matter of fact, choosing for example a zero OAM, thus fixed and symmetric slopes (as happens in figure 5a), the spatial profile can be modified by changing the number of periods of the triangle wave. In this situation a radially symmetric pattern can be generated characterized by a number of intensity lobes proportional to the period's number L.

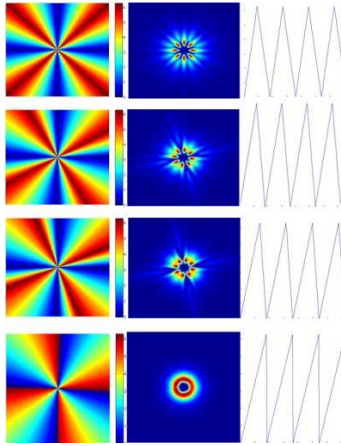

Figure 6. Triangular wave profile with different slopes and fixed number of periods. From left to right: imprinted phase profile, far field intensity of the reflected beam, steps' displacement. a) $m = 1$, OAM = 0; b) $m = 2$, OAM = 0.7; c) $m = 3$, OAM =1.3; d) $m = 128$, OAM = 4.

Figure 7 shows the intensity distributions obtained in this way; it can be observed that intensity profiles are different for even and odd numbers L. With fixed parity, higher order profiles, i.e. higher number of lobes, can be reached increasing L.

It is evident that even values of L correspond to intensity patterns made of 2L lobes while fields reflected by a mirror with an odd number of periods show a number of intensity maxima equal to L. The number of periods L and its parity can thus be considered as parameters able to control the field' spatial distribution in order, for example, to achieve a sort of spatial encoding.

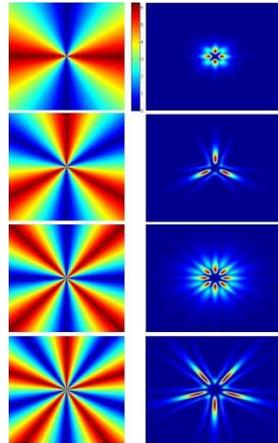

Figure 7. Phase (left column) and intensity profiles (right column) of triangular wave configurations with zero OAM for different values of the period's number L. Even periods generate a doubled number intensity lobes. a) L=2; b) L=3; c) L=4; d) L=5.

By programming the step's height it is possible to assemble many other configurations able to control both the topological charge, thus the orbital angular momentum, and the spatial profile. Since the reflection by the mirror acts on the incident beam only as a phase function of the kind expressed by Eq. (2), spatial profiles and OAM corresponding to a programmed height profile can be easily evaluated by means of Eqs. (4) and (5).

## 4. Experimental setup

Unlike OV generators like holograms or spiral phase plates which modify the transmitted beam, the proposed device works on reflection as happens for SLMs. Therefore a suitable experimental set up should be employed, like the one sketched in figure 8. A laser beam at the operating wavelength passes through a beam splitter and hits the mirror, which is controlled by a computer. A software allows to program the desired height profile, assigning a specified value for each step. This is then translated in an electrical signal which is transmitted to the actuators in order to displace the segments of the mirror.
The reflected beam impinges again on the beam splitter and is reflected toward the rest of the setup.

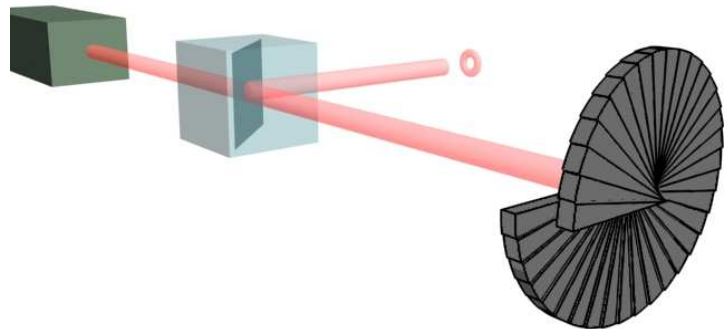

Figure 8. Experimental setup for the programmable mirror. The incident beam is reflected by the mirror and sent back to the beam splitter. The reshaped beam is then reflected by the beam splitter and collected.

## 5. Conclusions

A programmable segmented mirror able to generate optical vortex beams has been designed. This device is suitable to adjust OV's features like topological charge and can be used for different operating wavelengths. It was shown that the possibility of controlling each azimuthal step independently leads to an easier tuning of the optical vortex and, since a smaller displacement is required when using saw tooth height profiles, it allows to obtain higher values of the topological charge with respect to previously proposed devices. Moreover the sign of the OV can be reversed by simply inverting the slope of the ramps.
Furthermore it is possible to program an arbitrary height displacement; this means that it is possible to perform a continuous tuning of the OAM or to produce well controlled spatial field distributions which can be employed for spatial enconding.